\documentclass{iau}
\usepackage{graphicx}
\usepackage{bm}
\usepackage[usenames, dvipsnames]{color}

\usepackage{natbib}
\usepackage{url}
\usepackage{hyperref}
\hypersetup{
    final=true,
    pageanchor=true,
    colorlinks=true,
    breaklinks=true,
    linkcolor=blue,
    citecolor=blue,
    urlcolor=blue,
    pdfpagemode=UseNone,
    pdftitle={Full Stokes observations in the He I 1083 nm spectral region covering an M3.2 flare},
    pdfauthor={C. Kuckein, M. Collados, R. Manso Sainz, A. Asensio Ramos},
    pdfsubject={Solar Physics},
    pdfkeywords={Sun: flares, Sun: photosphere, Sun: chromosphere, Sun: magnetic
fields, techniques: polarimetric}}

\newcommand*\arcsec{\ensuremath{^{\prime\prime}}}

\title[Full Stokes observations of an M3.2 flare in the He\,{\scriptsize I} 1083 nm spectral region]
{Full Stokes observations in the He\,{\sc i} 1083 nm spectral region covering an M3.2 flare}

\author[C. Kuckein \etal]   
{Christoph Kuckein$^{1}$, Manuel Collados$^{2,3}$, Rafael Manso Sainz$^{2,3}$, and 
Andr\'es Asensio Ramos$^{2,3}$}

\affiliation{$^1$Leibniz-Institut f\"ur Astrophysik Potsdam (AIP), D-14482 Potsdam, Germany  \\
email: {\tt ckuckein@aip.de} \\
[\affilskip]
$^2$Instituto de Astrof\'\i sica de Canarias (IAC), E-38205 La Laguna, Tenerife,
Spain \\ [\affilskip]
$^3$Departamento de Astrof\'\i sica, Universidad de La Laguna, E-38206 La Laguna, Tenerife, 
Spain \\}

\pubyear{2015}
\volume{305}  
\jname{Polarimetry: From the Sun to Stars and Stellar Environments}
\editors{K.N. Nagendra, S. Bagnulo, \\ R. Centeno, \& M. Mart\'inez Gonz\'alez, eds.}

\begin{document}

\maketitle

\begin{abstract}
We present an exceptional data set acquired with the Vacuum Tower Telescope (Tenerife, Spain) covering
the pre-flare, flare, and post-flare stages of an M3.2 flare. The full Stokes spectropolarimetric observations 
were recorded with the Tenerife Infrared Polarimeter in the He\,{\sc i} 1083.0~nm spectral region. The object under
study was active region NOAA 11748 on 2013 May 17. During the flare the chomospheric He\,{\sc i} 1083.0~nm intensity 
goes strongly into emission. However, the nearby photospheric Si\,{\sc i} 1082.7~nm spectral line profile only gets 
shallower and stays in absorption. Linear polarization (Stokes $Q$ and $U$) is detected in all lines of the He\,{\sc i} 
triplet during the flare. Moreover, the circular polarization (Stokes $V$) is dominant during the flare, being 
the blue component of the He\,{\sc i} triplet much stronger than the red component, and both are stronger than 
the Si\,{\sc i} Stokes $V$ profile. The Si\,{\sc i} inversions reveal enormous changes of the photospheric magnetic 
field during the flare. Before the flare magnetic field concentrations of up to $\sim 1500$~G are inferred. During the 
flare the magnetic field strength globally decreases and in some cases it is even absent. After the flare the magnetic 
field recovers its strength and initial configuration. 

\keywords{Sun: flares, Sun: photosphere, Sun: chromosphere, Sun: magnetic
 fields, techniques: polarimetric}

\end{abstract}

\section{Introduction}
Solar flares are among the most dynamic and energetic events observed in the atmosphere of the Sun. 
The He\,{\sc i} 1083.0~nm spectral region offers a unique diagnostic tool
to study simultaneously chromospheric and photospheric magnetic fields. 
Intensity observations in the He\,{\sc i} 1083.0~nm triplet have been used in the past
to study flares at the limb \citep[e.g.,][]{you92}, and on the disk 
\citep[e.g.,][]{malanushenko99,du08,li07}, 
and the formation of the intensity profile theoretically considered 
\citep[e.g.,][]{ding05}.
Circular polarization measurements in flares have been carried out by 
\citet[][]{penn95}.
More recently, several works using the He\,{\sc i} triplet in flares have appeared.
\citet[][]{zeng14} have presented high resolution filtergrams of a C-class 
flare to study the mechanism which leads to He\,{\sc i} 1083.0~nm emission. 
\citet[][]{akimov14} have described an X1.4 flare as observed
with a spectroheliograph in the vicinity of He\,{\sc i} 1083.0~nm. 
Using spectropolarimetry \citet[][]{sasso11,sasso14} have covered
a C2.0 flare with one spectral scan and focused their analysis on the magnetic structure of an activated filament.
Finally, \citet[][]{judge14} have presented observations that cover the evolution of an X1.0 flare. 
These authors concentrate on several aspects of the flare, among others, the photospheric magnetic field before 
and after the flare. 

\newpage
The full Stokes observations in the He\,{\sc i} 1083.0~nm spectral region presented in this work cover the pre-flare, 
flare, and post-flare stages of an M3.2 flare. Fortunately, the slit was at the right time and position, 
covering the whole evolution of the flare. 
The first results concerning the evolution of the photospheric magnetic field 
have been recently published by \citet[][]{kuckein15}.

\begin{figure}[t]
\begin{center}
\includegraphics[scale=0.45]{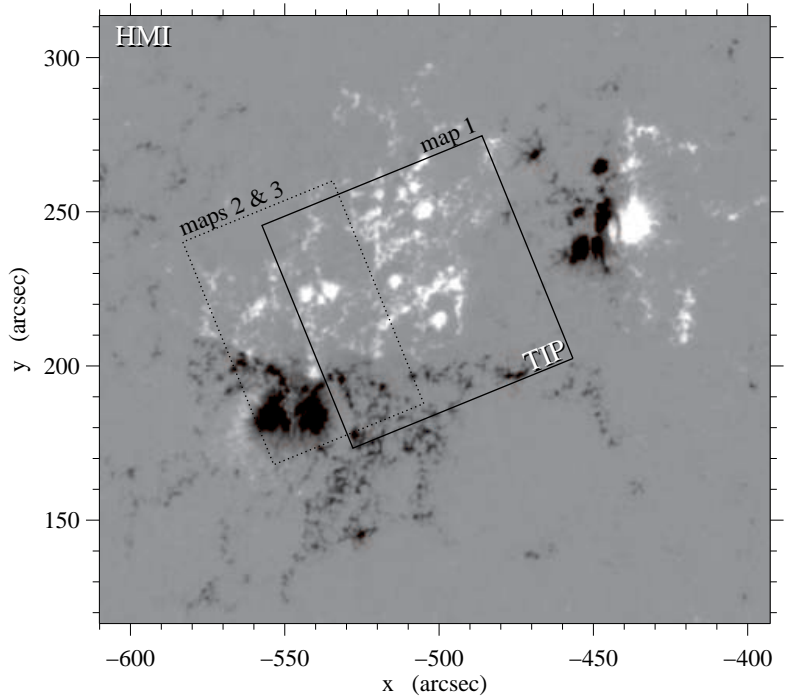}
\caption{HMI line of sight (LOS) magnetogram of active region NOAA 11748 on 2013 May 17 at 08:57~UT. The magnetogram is 
clipped between $\pm 700$~G. A filament was lying on top of the PIL. 
The solid and dashed boxes show the approximate scanned
area of TIP-II. Solar north and west correspond to up and right.  }
\label{fig1}
\end{center}
\end{figure}

\section{Observations}
Active region NOAA 11748 produced 14 flares, between classes C and X. On 2013 May 17, an M3.2 flare
was captured with the telescope. The flare started at 8:43~UT and finished at 9:19~UT. The peak of the flare occurred 
at 8:57~UT. The ground-based observations were carried out with the Tenerife Infrared Polarimeter
\citep[TIP-II;][]{tip2} attached to the Echelle spectrograph of the Vacuum Tower Telescope 
\citep[VTT; Tenerife,][]{vonder98}. 
The main target was a filament lying on top of the polarity inversion line (PIL) at 
coordinates (N~11$^\circ$, E~36$^\circ$). 
The first scan with the slit, from 7:48 to 8:36~UT, was the largest scan and covered 
77\arcsec\ in the scan direction. This map covered the pre-flare stage. The second map, from 8:36 to 
9:06~UT, was smaller and covered 52.5\arcsec\ in the scan direction. In this map the slit passed over the flare at its 
peak ($\sim$ 
8:57~UT). 
The third map, from 9:06 to 9:37~UT, covered the same spatial area as the previous
map and included the post-flare phase of the flare. All three scanned regions are superimposed in Fig. \ref{fig1} on a 
context magnetogram of the Helioseismic and Magnetic Imager \citep[HMI;][]{HMI} on board the Solar 
Dynamics Observatory \citep[SDO;][]{SDO}.

The spectral range included full Stokes spectropolarimetry in the He\,{\sc i} 1083.0~nm vicinity. The 
spectra contained the photospheric Si\,{\sc i} line at 1082.7~nm, the chromospheric He\,{\sc i} triplet, which 
comprises three spectral lines (the ``blue'' component at 1082.9~nm and the blended ``red''
component at 1083.0~nm), and two telluric lines.  The spectral sampling was $\sim 11.1$~m\AA/pixel.
The exposure time per slit position was 10~s and the scanning step 0.35\arcsec. The pixel size along the slit was 
0.17\arcsec. The Kiepenheuer-Institute Adaptive Optics System \citep[KAOS;][]{berkefeld10} was 
locked on small structures, like pores and small penumbrae, and significantly improved the image quality under
fair seeing conditions. 

Flare processes are complex phenomena which involve many different layers of the solar atmosphere. In
this study, we simultaneously compare two different layers. 
The formation height of the He\,{\sc i} triplet corresponds to the upper
chromosphere \citep[][]{avrett94} while the photosphere is represented by the 
Si \,{\sc i} line \citep[e.g.,][]{bard08}.

\section{Data reduction}
All data sets were corrected for flat-field and dark frames. Moreover, the standard polarimetric calibration for 
the TIP-II instrument was carried out \citep[][]{collados99,collados03}. 

For the normalization procedure of the Stokes profiles we computed a continuum value for each position of the slit. 
This was needed because the intensity level changed throughout the scans because of changing air mass and center-to-limb
variations. For this purpose we used all available maps from that day (six maps). In each map, several 
scans were averaged along the slit over a quiet Sun area. Afterwards, a second-order least-square polynomial fit 
was carried out to retrieve the quiet Sun continuum values individually for each scan step.
All pixels along the slit within one scan step were divided by their corresponding quiet Sun continuum value. 

The wavelength calibration was carried out using the two nearby-lying telluric lines. In an averaged quiet Sun area, we 
calculated the distance between both telluric lines and compared it with the distance separating the same 
telluric lines in the Fourier Transform Spectrometer spectrum \citep[FTS;][]{neckel84}. 
The division of both distances yielded the spectral sampling. We followed the procedure explained in
Appendices A and B of \citet[][]{kuckein12b} to retrieve the wavelength axis on an
absolute scale, i.e., corrected for Earth's orbital motions, solar rotation, and the solar gravitational redshift.

\begin{figure}[t]
\begin{center}
\includegraphics[width=\textwidth]{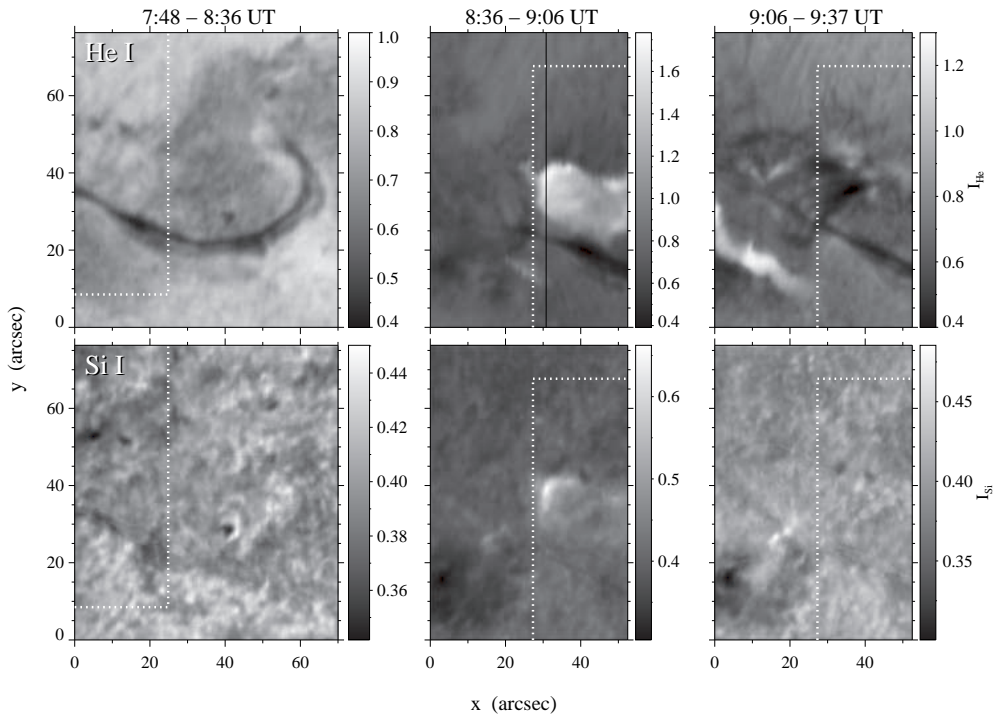}
\caption{Monochromatic slit-reconstructed images of the red component of He\,{\sc i} (upper row)
 and the Si\,{\sc i} line core (lower row). The different stages of the flare are shown
from left to right: pre-flare, flare and post-flare maps. The dashed boxes indicate the common area between all 
three maps. The overlapping area can also be identified in Fig. \ref{fig1}. The vertical solid line in the 
middle map shows the location of the Stokes profiles in Fig. \ref{fig3}.  }
\label{fig2}
\end{center}
\end{figure}

\section{Results and discussion}
\subsection{Analysis of the Stokes profiles}

The data sets in Fig. \ref{fig2} are represented as monochromatic slit-reconstructed intensity images. The upper row 
shows the
line core of the He\,{\sc i} red component and the time range of the scans. The three panels correspond to the 
pre-flare, flare, and post-flare scans,
respectively. In the first scan, we focus on the image inside the dashed box, which outlines the common part of the FOV 
between all three maps. A filament is lying on top of the PIL. In addition, two pores 
are seen above
the filament at around $y \sim 53$\arcsec. The middle panel shows the flare phase. 
Strong He\,{\sc i} 1083.0~nm emission can
be identified between the filament and the pores. 
Turning to the third panel, which is aligned with the middle panel, the presence of post-flare loops is clear. 
These loops must have formed very quickly ($< 30$~min). There is still some emission of He\,{\sc i} in the 
post-flare map, but weaker (see the colorbar) and in a different area of the FOV. 

The lower three panels in Fig. \ref{fig2} show the pre-flare, flare, and post-flare maps as seen in the 
normalized Si\,{\sc i} line core intensity. 
Although the line core gets shallower during the flare (see the colorbar in the middle panel), by far it never goes 
into emission. 

\begin{figure}[t]
\begin{center}
\includegraphics[width=\textwidth]{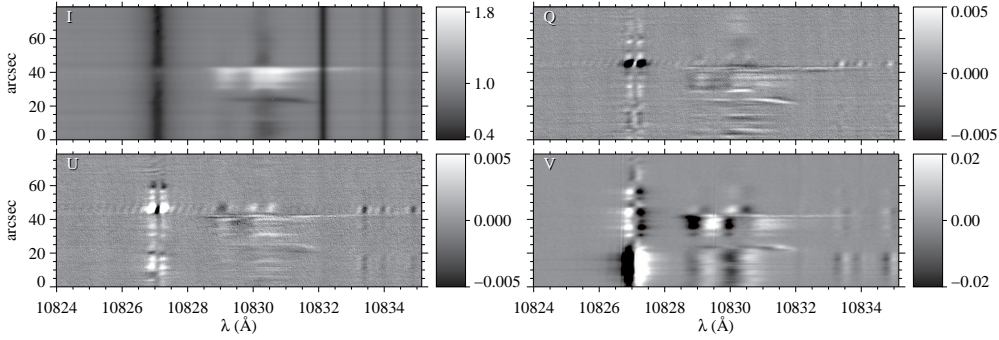}
\caption{Starting top left and moving clockwise: normalized Stokes profiles $I$, $Q$, $V$, 
and $U$. The profiles correspond to the vertical solid line in the top-middle panel in Fig. \ref{fig2}. 
From left to right the 
following spectral lines can be seen: 
Si\,{\sc i}, the He\,{\sc i} triplet, and two telluric lines. 
Stokes $Q$ and $U$ are clipped at $\pm 0.005$ and 
Stokes $V$ at $\pm 0.02$.}
\label{fig3}
\end{center}
\end{figure}

Examples of Stokes profiles during the flare are shown in Fig. \ref{fig3}. Starting in the upper left corner and 
counting clockwise: Stokes $I$, $Q$, $V$, and $U$. 
Note that the normalized Stokes parameters are given on different scales for each panel. 
The profiles correspond to the vertical solid 
line shown in the upper middle panel of Fig. \ref{fig2}.
The Stokes $I$ panel shows a broad vertical line on the lefthand which corresponds to the Si\,{\sc i} line. The next 
two lines to the right correspond to the He\,{\sc i} triplet. The single most striking observation from this panel 
is the strong emission of the He\,{\sc i} triplet, with an intensity ratio to the continuum of up to $\sim 1.86$. 
Some emission profiles seem to be strongly redshifted as they almost reach the second telluric line. Similarly,
redshifts occur in the absorption profiles which correspond to the filament at $y \sim 24$\arcsec. 

The linear polarization, Stokes $Q$ and $U$ panels, are clipped between $\pm 0.005$. Especially in the area were 
the flare occurs (He\,{\sc i} emission, $I > 1$), one-lobe polarization signatures are found that cannot be ascribed to 
the typical Zeeman shapes. Interestingly, both the blue and red components of the He\,{\sc i} triplet show 
linear
polarization signatures of the same sign. As for the Stokes $V$ panel (saturated at $\pm 0.02$), there is significant 
circular polarization seen in the He\,{\sc i} triplet during the flare. Curiously, the He\,{\sc i} Stokes $V$ profile 
is reversed with respect to the Si\,{\sc i} one where the flare occurs. This happens because only He\,{\sc 
i} goes into emission during the flare. 
Additionally, during the flare the Stokes $V$ amplitude is larger in He\,{\sc i} 
as compared to Si\,{\sc i}. Furthermore, the He\,{\sc i} Stokes $V$ amplitude of the blue component is
significantly larger than the Stokes $V$ amplitude of the red component. The profile of the red 
lobe of Stokes $V$ from the red component of He\,{\sc i} is broad and somehow displaced and stretched towards bigger 
wavelengths (redshifted). This is not the case for the blue component nor for the Si\,{\sc i} Stokes $V$ profiles.

\subsection{Photospheric magnetic field}
We carried out full Stokes inversions of the photospheric Si\,{\sc i} line using \textit{Stokes Inversions based on
Response functions} \citep[SIR;][]{SIR}. The model atmosphere covered a range
of the logarithm of the LOS continuum optical depth at 500~nm between $1.4 \leq \log \tau \leq -4.0$. To infer the 
magnetic field at the height of granulation, we will focus on a constant height at $\log \tau = -1$. 
Furthermore, we will concentrate on a region of interest within the FOV which only includes the 
flare. The region of interest fulfills the criterion that He\,{\sc i} goes into emission ($I > 1$, 
see top-middle panel of Fig. \ref{fig2}). In other words, the region of interest is inside the rectangle with 
coordinates $x \sim [30\arcsec,50\arcsec]$ and $y \sim [25\arcsec,40\arcsec]$ in the middle panel of Fig. \ref{fig2}. 
The same area 
was analyzed in the pre-flare and post-flare maps. After carefully comparing the inferred photospheric magnetic field 
before, during, and after the flare we found the following: (1) Concentrations of several seconds of arc of strong 
magnetic 
fields ($< 1500$~G) are seen before the flare. (2) During the flare, the field strength globally decreased and even 
vanished in some areas. (3) After the flare, the magnetic field has partially recovered its strength and pattern. 
\citet[][]{kuckein15} presented a detailed map of the photospheric magnetic field and the associated 
Doppler velocities. 

We conclude that remarkable changes of the photospheric magnetic field of up to 1500~G are found during this M3.2 
flare. These changes happen in short time periods ($\leq 30$~min). While changes of the photospheric
magnetic field during flares  have been reported in the past \citep[e.g.,][]{petrie10}, such a 
strong decrease was not seen before.

A natural progression of this work is to invert the He\,{\sc i} triplet and infer the 
chromospheric magnetic field and Doppler velocities. In terms of directions for future research, further 
multi-height and multi-wavelength observations of the evolution of flares are crucial. A promising telescope
for carrying out such observations is the 1.5-meter GREGOR telescope \citep[][]{gregor} located 
at the Observatorio del Teide in Tenerife. The combination of its two principal instruments,
the GREGOR Infrared Spectrograph \citep[GRIS;][]{collados12} equipped with TIP-II 
and the GREGOR Fabry-P\'erot Interferometer \citep[GFPI;][]{puschmann12} will
simultaneously cover several heights in the solar atmosphere. Integrating sensitive near-infrared 
polarimetry with high spatial and temporal resolution observations exploits the 
synergy between scanning spectrographs and imaging spectropolarimeters for flare research.

\begin{acknowledgements}
\noindent CK greatly acknowledges the SOC of the IAUS305 for the travel support received from the HAO of the National 
Center for Atmospheric Research (NCAR) in USA. The VTT is operated by the Kiepenheuer-Institute for Solar Physics in 
Freiburg, Germany, at the Spanish Observatorio del Teide, Tenerife, Canary Islands. The authors would like to thank C. 
Denker for carefully reading the manuscript. 
\end{acknowledgements}

\end{document}